\documentclass[aip,cha,reprint]{revtex4-1}

\usepackage{bm}
\usepackage{color}
\usepackage{amsmath,amssymb}
\usepackage{graphicx}

\begin{document}

\title{Bifurcation in the angular velocity of a circular disk propelled by symmetrically distributed camphor pills}

\author{Yuki~Koyano}
\email{y.koyano@chiba-u.jp.}
\affiliation{Department of Physics, Chiba University, Chiba 263-8522, Japan}

\author{Hiroyuki~Kitahata}
\affiliation{Department of Physics, Chiba University, Chiba 263-8522, Japan}

\author{Marian~Gryciuk}
\affiliation{Institute of Physical Chemistry, Polish Academy of Sciences, Warsaw 01-224, Poland}

\author{Nadejda~Akulich}
\affiliation{Department of Chemistry, Technology of Electrochemical Production and Electronic Engineering Materials, 
Belarusian State Technological University, Minsk 220006, Belarus}

\author{Agnieszka~Gorecka}
\affiliation{School of Physics and
Astronomy, Monash University, Clayton, Victoria 3800, Australia}

\author{Maciej~Malecki}
\affiliation{Institute of Physical Chemistry, Polish Academy of Sciences, Warsaw 01-224, Poland}

\author{Jerzy~Gorecki}
\affiliation{Institute of Physical Chemistry, Polish Academy of Sciences, Warsaw 01-224, Poland}

\date{\today}

\begin{abstract}
We studied rotation of a disk propelled by a number of camphor pills symmetrically distributed at its edge.
The disk was put on a water surface so that it could rotate around a vertical axis located at the disk center.
In such a system, the driving torque originates from surface tension difference resulting from inhomogeneous surface concentration of camphor molecules released from the pills.
Here we investigated the dependence of the stationary angular velocity on the disk radius and on the number of pills.
The work extends our previous study on a linear rotor propelled by two camphor pills [Phys. Rev. E, 96, 012609 (2017)].
It was observed that the angular velocity dropped to zero after a critical number of pills was exceeded.
Such behavior was confirmed by a numerical model of time evolution of the rotor.
The model predicts that, for a fixed friction coefficient, the speed of pills can be accurately represented by a function of the linear number density of pills. We also present bifurcation analysis of the conditions at which the transition between a standing and a rotating disk appears.
\end{abstract}


\maketitle

{\bf
Camphor is one of many substances that form a layer on the water surface and modify the surface characteristics.
The presence of camphor molecules at water surface reduces its surface tension.
The camphor surface concentration profile results from the balance between camphor release from the source, its transport, evaporation and dissolution.
Inhomogeneities in camphor concentration around a floating object activate the motion because the object is propelled towards the region characterized by the lowest concentration.
In this paper, we study rotation of a disk propelled by a number of camphor pills.
The disk angular velocity nonlinearly depends on the number of pills and falls to zero when the number of pills exceeds the critical value. 
The developed model treats the time evolution of camphor surface concentration as a reaction-diffusion process.
It reproduces qualitatively experimental observations, which confirms model usefulness for simulations of systems with surface interactions.
Moreover, the model can be reduced and allows for an analytical investigation of bifurcation between the standing and the rotating states of a disk.
We believe that our results are important because they describe a realistic complex system for which a bifurcation can be investigated analytically.}

\section{Introduction}

Studies on self-propelled objects have become popular in the recent years because the behavior of many such systems shows similar characters of motion to that expressed by living organisms.
Self-propelled motion can be observed in systems with embedded asymmetry of system structure and interactions.
For example, Janus particles, characterized by different rates of reactions at different parts of their surface, can move in the direction determined by the chemical activity\cite{Janus,Janus2}.
There are also objects in which the boundaries direct a jet of reaction products and force the motion\cite{rumunka}.
The self-propelled motion can be also observed for symmetric objects in which the symmetry is broken by processes that generate the motion.
Such systems include droplets where Belousov-Zhabotinsky (BZ) reaction proceeds\cite{kitahata_JCP}.
The interfacial tension between a droplet and the surrounding oil phase is related to the level of catalyst oxydization\cite{yoshikawa-old}.
If a droplet is sufficiently large then homogeneous oscillations are observed and for yet larger droplets, a propagating excitation pulses can appear\cite{loc-2016}.
The related changes in interfacial tension generate a jump of the droplet in the direction of pulse propagation\cite{kitahata-BZ-motion,kitahata-BZ-motion2}.
However, since the direction of an excitation pulse is random, a symmetric BZ droplet can be shifted in a stochastic direction and there are no factors that can stabilize the direction of motion.

In this paper, we are concerned with self-propelled motion induced by interfacial phenomena related to dynamically changing surface concentration of camphor molecules.
It is known that if a piece of camphor is placed on the water surface then camphor molecules hardly dissolve in water, but the majority of them forms a layer on the water surface\cite{soh,pccp,19,camphor,7,8}.
In typical experimental conditions, this layer is unstable, because camphor molecules continuously evaporate.
The water surface tension decreases as camphor surface concentration increases\cite{soh,suematsu,Karasawa}.
As a consequence, the force acting on a camphor piece is directed towards the neighboring region with the lowest camphor surface concentration.
One of the simplest and most known camphor-propelled objects is a camphor boat i.e., a boat-shaped piece of plastic with a bit of camphor glued at its stern\cite{Kohira,Shimokawa}.
Such configuration of camphor-propelled objects breaks system symmetry.
The surface concentration of released camphor molecules around the stern is higher than that around the bow, which decreases the surface tension in the stern area.
As a result, the boat moves forward.

Geometrically symmetric objects can be also propelled by camphor pieces, because there is a positive feedback between the generated force (or torque) and the direction of object motion.
Let us consider a camphor disk reclining on the water surface.
It releases camphor molecules around, but both formation of a camphor layer and the evaporation of camphor molecules are subject to fluctuations.
If an area characterized by a low surface concentration of camphor appears close to the disk, then the disk is shifted towards the area because it is attracted by the region with higher interfacial tension.
When the disk is shifted from the original position, the surface camphor concentration in front of the disk is lower than in the region behind the disk, because the area in front of the disk has been more distant from the camphor source than the region behind the disk.
Therefore, the disk motion continues up to the moment the disk hits the boundary or it is repelled by water meniscus near the boundary.

Studies on self-propelled rotational motion are interesting since such motion occurs in a confined space, thus effect of boundaries can be neglected.
There have been several reports on systems that show spontaneous rotation including systems with broken chiral symmetry\cite{Leonardo,Takinoue,Hayakawa,Lowen,Yamamoto,Mitsumata,Frenkel,Bazant} and systems in which rotation occurs through the spontaneous breaking of chiral symmetry \cite{Pimienta,Pimienta2,Takabatake,Nagai,Takabatake3,koyano_PhysicaD,Bassik,Ebata}.
For camphor driven systems, it has been demonstrated \cite{single_rotor,hexagon} that fluctuations of surface camphor concentration can induce initial rotation that is supported by the positive feedback between the direction of motion and the concentration gradient, like for the translational motion of a camphor disk mentioned above.

However, there have not been too many studies in which mathematical modeling of self-propelled rotational motion has been compared with experimental results. 
In this respect, systems that are propelled by camphor pieces are worth considering because they can be analyzed using a simplified model of their time evolution.
This model is based on a reaction-diffusion equation for camphor surface concentration coupled with the Newtonian equation of motion for the camphor pieces\cite{pccp,IKN,single_rotor,doi,Kitahata_PhysicaD,20,2d}.

In our previous paper\cite{single_rotor}, we considered a camphor rotor with two camphor pills at the ends of a plastic stripe. 
The pills were floating on the water surface, whereas the stripe was elevated above the surface.
The system was allowed to rotate around a vertical axis at the center of the stripe. 
We observed that such rotor can move only after a distance between the camphor pills was larger than the critical one.
For this system, the rotor radius can be considered as a bifurcation parameter.
The mathematical model of the spontaneous symmetry breaking can be formulated in terms of pitchfork bifurcation in dynamical systems.
Here, as a generalization of previously studied problem, we consider disk-shaped rotors powered by a number of camphor pills.
The pills are symmetrically distributed at the disk edge.
There are two parameters that describe the system: the disk radius and the number of pills.
It can be expected that a disk propelled by greater number of pills rotates at a higher angular velocity.
On the other hand, a disk with many camphor pills seems equivalent to a disk with a continuous camphor source along its edge, which clearly does not rotate.
Therefore, there is a question on how the angular velocity of a disk depends on the number of camphor pills.
We have performed experiments and analyzed the stationary angular velocity of the disk as a function of the both parameters.
The results are reported in Section II.
It turns out that, for small disk radius, the angular velocity drops to zero when the number of pills is large.
If the disk is large, then the angular velocity weakly depends on the number of pills attached in the same range of pill numbers.
In Section III, we present a mathematical model describing the disk rotation.
Results of numerical simulations presented in Section IV allowed us to determine the values of model parameters for which the qualitative agreement with experimental results is obtained.
Section V is concerned with the analytical methods used to study disk evolution and with the analysis of bifurcation between the rotating and the still disk.
Finally in the Section VI, we present numerical arguments that, for a fixed friction coefficient, the speed of pills can be accurately represented by a function of a single argument: the linear number density of pills.
We demonstrate that such behavior can be found in experimental results.

\section{Experiments}

\begin{figure}
	\begin{center}
		\includegraphics{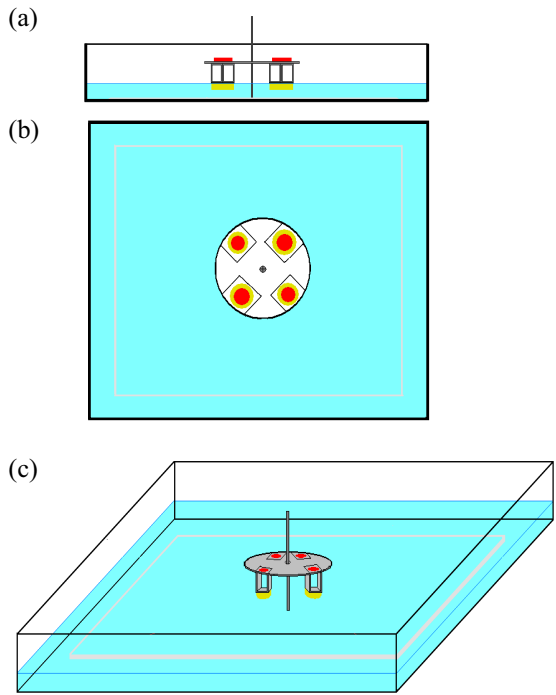}
		\caption{Disk with pills glued at the bottom of supporting columns studied in experiments: (a) the side view, (b) the top view and (c) the slanted view of the disk propelled by 4 pills (the pills are marked yellow).
The disk could rotate around the vertical axis fixed at the tank center.
The columns were made of bended plastic stripe.
}
		\label{fig1}
	\end{center}
\end{figure}

We study the angular velocity of a disk propelled by a number of camphor pills located on water surface.
The system is illustrated in Fig.~\ref{fig1}.
The disk could rotate around a vertical axis located at its center.
The pills were symmetrically distributed close to the disk edge.
They were glued to the columns located below the disks such that the pills were in contact with water, whereas the disk was elevated over the water surface to avoid generation of an extensive hydrodynamic flows by the moving disk~\cite{hexagon}.
The camphor surface concentration on water is determined by a number of physicochemical processes such as the inflow of camphor molecules from the pills, diffusion at the surface, evaporation into the air, and dissolution into the water~\cite{pccp}.
The randomness embedded in these processes can lead to a nonuniform surface tension generating the driving torque of the disk.
Like in the camphor disk motion discussed in Introduction, there is a positive feedback between the disk angular shift and the force generating the torque.

\begin{figure}
	\begin{center}
		\includegraphics{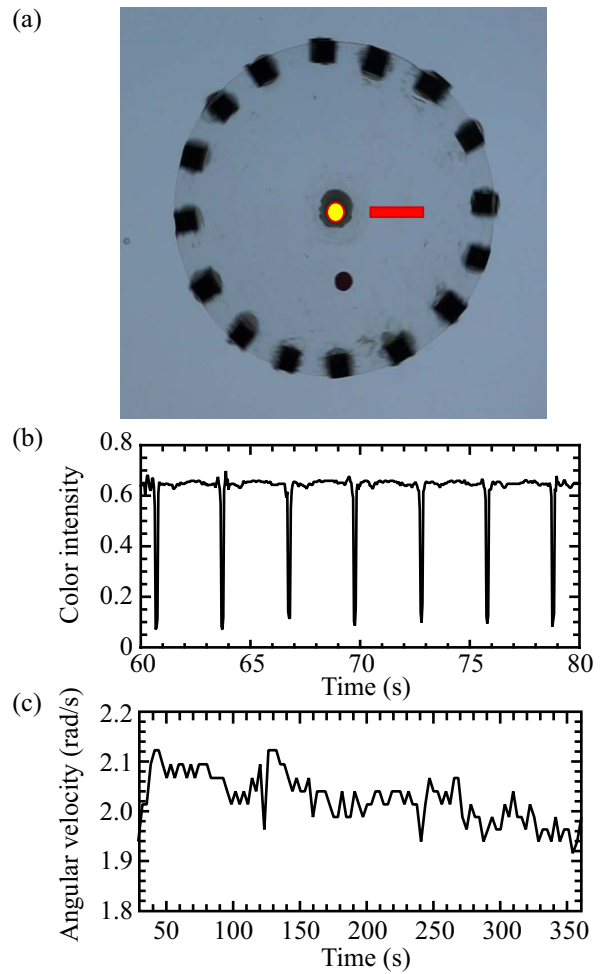}
		\caption{Illustration of the experimental data processing.
(a) A disk ($R=2$~cm, $N=16$) seen from above.
A black marker glued on the disk was used to measure the angular velocity.
(b) Time dependent color intensity at the center of red line (cf. (a)) during the disk rotation read from the frames of filmed experiment.
The minima correspond to moments when the marker crossed the line.
(c) Angular velocity of the disk as a function of time.}
		\label{fig2}
	\end{center}
\end{figure}

We investigated the stationary angular velocity depending on the number of pills $N$ and on the distance between the axis and the pill center $\ell$.

The pills of mass $m$ and radius $\rho$ were tangent to the disk, so the disk radius $R =\ell + \rho$. In the experiments, we used $150$~ml of water poured into a square tank (tank side $12$~cm) so that the water level was $1$~cm. 
Water was purified using a Millipore system (Elix 5) and its temperature was $ 22 \pm 1$ ${}^\circ$C.
In the experiments, we used commercially available camphor ($ 99 \%$ purity, Sigma-Aldrich) without further purification.
The pills were made by pressing camphor in a pill maker. The camphor pill radius was $\rho = 0.15$~cm and the height was 0.1~cm.

\begin{figure}
	\begin{center}
		\includegraphics{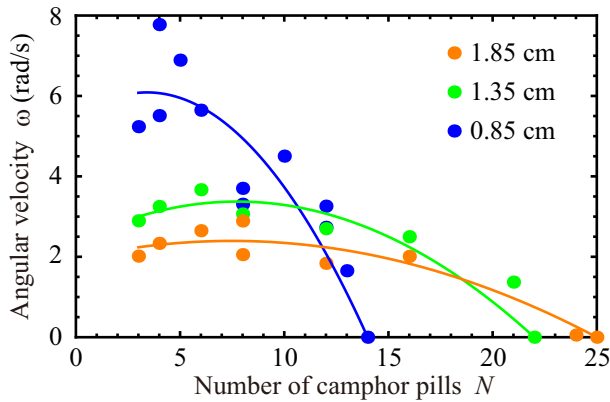}
		\caption{Experimental results on the angular velocity $\omega(\ell,N)$ as a function of the number of camphor pills $N$ for 3 selected values of $\ell$ -- the distance between disk axis and the dot center (0.85, 1.35 and 1.85~cm).
		Dots represent experimental data and curves show their fit using a quadratic polynomial with zeros at $N= 14$, $22$ and $25$, respectively.
		Blue, green, and orange points correspond to $\ell =0.85$, $1.35$, and $1.85$~cm, respectively.}
		\label{fig_exp_vel}
	\end{center}
\end{figure}

Figure~\ref{fig2} illustrates the analysis of experimental data.
A marker was attached on the disk surface.
Its position was recorded on a movie and analyzed using ImageJ program\cite{ij}.
We applied two methods to obtain the angular velocity.
In one of them, we calculated the angular velocity as a function of time using positions of the marker in the consecutive frames of the movie. 
In the second method, we analyzed time-dependent color intensity in the region marked as a red line in Fig.~\ref{fig2} (a).
We observed a minimum in color intensity every time the marker passed this region.
The angular velocity averaged over a single rotation was calculated from the time difference corresponding to the successive minima.
Both methods gave similar results.
In a typical experiment, the angular velocity was quite stable (cf. Fig.~\ref{fig2} (c)) and weakly depended on time at the time scale of a few minutes, which corresponded to over $100$ rotations in the presented case.
A small decrease in angular velocity with time can be related to the increase in camphor concentration in bulk aqueous phase.
There is a slow dissolution of camphor into the bulk aqueous phase.
The dissolved molecules migrate to the surface and contribute to the surface concentration of camphor.
The contribution is homogeneous and reduces gradient of camphor surface concentration resulting from local release and evaporation.
As a result, the inhomogeneities in surface tension are decreased and so is the torque.

Figure~\ref{fig_exp_vel} summarizes the experimental results obtained for 3 different disk radii.
It shows the average angular velocity $\omega(\ell,N)$ measured within first 6 minutes of rotation.
For a small number of pills ($N \le 10$), the angular velocity for fixed $N$ was a decreasing function of $\ell$.
Considering the dependence of $\omega(\ell,N)$ on $N$, we observed that for small disk radii the angular velocity rapidly decreased with the increasing number of pills.
The disk of $\ell=0.85$~cm powered by $N=14$ pills randomly moved in both directions, but it did not show unidirectional rotation lasting more than a second.
For a larger disk radii ($\ell=1.35$~cm and $\ell=1.85$~cm), the angular velocity was slowly decreasing with $N$ in the range of $5 < N < 20$.
For $\ell=1.35$~cm and $N=21$, the angular velocity was around the half of the value observed at small $N$.
The disk did not rotate when $N \ge 22$.
In the case of $\ell=1.85$~cm, we observed a stable rotation with a very long period of $115$~s for $N=24$ and the disk stopped at $N=25$. 
In all experimental results $\omega(N=3) < \omega(N=4)$, so we cannot exclude a maximum of angular velocity at small $N \in \{4,5\}$. The lines in Figure~\ref{fig_exp_vel} show a fit using the second order polynomials with zeros at the smallest number of pills for which the disk was not rotating.

\section{Model}

In this section, we introduce a mathematical model for a disk propelled with $N$ camphor pills attached.
We set the coordinates so that the center of the disk, i.e., the rotation axis, is located at the origin.
The motion of the disk is described by the characteristic angle $\phi (t)$.
All $N$ camphor pills are located at the distance of $\ell$ from the origin and have equal spacing.
Thus the position of $j$-th camphor pill, $\bm{\ell}_j(t)$, can be described as:
\begin{equation}
\bm{\ell}_j(t) = \ell \left[ \cos \left(\phi(t) + \frac{2 \pi j}{N} \right) \bm{e}_x + \sin \left(\phi(t) + \frac{2 \pi j}{N} \right) \bm{e}_y \right], \label{position}
\end{equation}
for $j = 0, \cdots, N-1$, where $\bm{e}_x$ and $\bm{e}_y$ denote the unit vectors in $x$- and $y$-directions, respectively.

The camphor pills release camphor molecules to the water surface, and the camphor molecules diffuse at the water surface.
Some camphor molecules sublimate to the air and some dissolve to the water bulk phase.
All these processes are taken into account in the equation for the time evolution of the camphor surface concentration:
\begin{equation}
\frac{\partial c(\bm{r},t)}{\partial t} = \nabla^2 c(\bm{r},t) - c(\bm{r},t) + \sum_{j=0}^{N-1} f(\bm{r}; \bm{\ell}_j), \label{diffusion}
\end{equation}
where $c(\bm{r},t)$ is the surface concentration of camphor molecules at the position $\bm{r}$ and time $t$.
The first term corresponds to the diffusion, and the second one to sublimation to the air and dissolution to the water bulk phase.
The last term $f(\bm{r}; \bm{\ell}_j(t))$ denotes the release of camphor molecules from the $j$-th pill, which is represented as 
\begin{equation}
f(\bm{r}, \bm{\ell}_j(t)) = \frac{1}{\pi \rho^2} \Theta\left( \rho - \left| \bm{r} - \bm{\ell}_j(t) \right| \right), \label{supply}
\end{equation}
where $\rho$ corresponds to the camphor pill radius and $\Theta(\xi)$ is the Heaviside's step function, i.e., $\Theta(\xi) = 1$ for $\xi \geq 0$ and $\Theta(\xi) = 0$ for $\xi < 0$.

The camphor molecules reduce the surface tension of the water surface, and the camphor pills are driven by the surface tension around it, which induces the spinning motion of the disk.
The dynamics on this motion is described as
\begin{equation}
I \frac{d^2 \phi}{dt^2} = - \eta_r \frac{d \phi}{dt} + \mathcal{T}.
\end{equation}
Here, $I$ is the momentum of inertia of the disk, and it is approximated using the mass of one camphor pill and its supporting column $m$ as
\begin{equation}
I = N m \ell^2 = \pi N \sigma \rho^2 \ell^2,
\end{equation}
where $\sigma$ is the average surface density.
$-\eta_r d\phi/dt$ is the torque originating from the friction force working on the camphor pills and $\eta_r$ is described using the friction coefficient per unit area, $\kappa$,
\begin{equation}
\eta_r = \pi N \kappa \rho^2 \ell^2, \label{friction}
\end{equation}
as is derived in the previous paper\cite{single_rotor,footnote1}.
$\mathcal{T}$ is the torque exerting on the disk, which is represented as
\begin{align}
\mathcal{T} =& \sum_{j=0}^{N-1} \bm{\ell}_j \times \bm{F}_j,
\end{align}
where
$\bm{F}_j$ is the driving force of the $j$-th camphor pill induced by the surface tension gradient.
It should be noted that constraint force should work on each pill to maintain the composition of the disk, but the direction of the constraint force working on the $j$-th pill is the same as $\bm{\ell}_j$, and therefore it does not affect the torque.
$\bm{F}_j$ is described as
\begin{equation}
\bm{F}_j = \int_0^{2\pi} \gamma \left(c \left( \bm{\ell}_j + \rho \bm{e}(\theta) \right) \right) \bm{e}(\theta) \rho d\theta,
\end{equation}
where $\gamma(c)$ is the surface tension depending on the camphor surface concentration, and $\bm{e}(\theta)$ is a unit vector in the direction of $\theta$, i.e., $\bm{e}(\theta) = \cos \theta \bm{e}_x+ \sin \theta \bm{e}_y $.
For simplicity, we set
\begin{equation}
\gamma(c(\bm{r},t)) = \gamma_0 - k c(\bm{r},t), \label{surfacetension}
\end{equation}
where $\gamma_0$ is water surface tension, and $k$ is a positive constant.
Hereafter, we set $k = 1$.

Taken in all, we obtain
\begin{equation}
\pi N \sigma \rho^2 \ell^2 \frac{d^2\phi}{dt^2} = - \pi N \kappa \rho^2 \ell^2 \frac{d\phi}{dt} + \mathcal{T},
\end{equation}
or
\begin{equation}
\sigma \frac{d^2\phi}{dt^2} = - \kappa \frac{d\phi}{dt} + \frac{1}{\pi N \rho^2 \ell^2} \mathcal{T}. \label{eqmotion}
\end{equation}

All variables in the equations above are dimensionless.
Let us assume that $D$ is the diffusion constant of camphor molecules, $a$ is the combined rate of sublimation and dissolution, and $f$ is the release rate of camphor molecules from one camphor pill.
The dimensionless variables are defined such that the time unit is the characteristic time of combined sublimation and dissolution $1/a$, the length unit is the diffusion length $\sqrt{D / a}$, and the concentration unit is the ratio $f/a$.
Our approach approximately treats the hydrodynamic effects.
Due to surface tension gradients, the Marangoni flow should appear in the system.
In our approach, $D$ is an ``effective'' diffusion constant that allows to include the camphor transport related to the Marangoni effect~\cite{JCP,suematsu}.

\section{Numerical simulations}

\begin{figure}
\begin{center}
\includegraphics{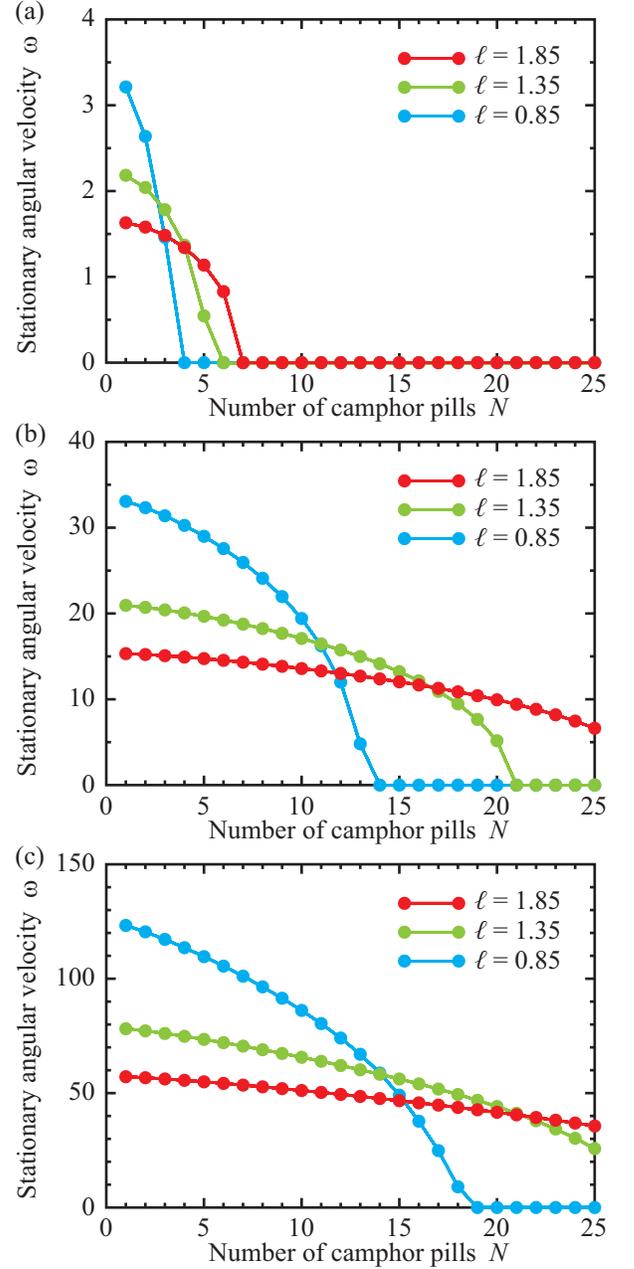}
\end{center}
\caption{Results of numerical calculation on the stationary angular velocity, $\omega(\ell, N)$, expressed in the dimensionless units, depending on the number of camphor pills, $N$, for each $\ell$.
Cyan, green, and red plots correspond to $\ell =0.85$, $1.35$, and $1.85$, respectively.
(a) $\kappa = 0.1$, (b) $\kappa = 0.01$, and (c) $\kappa = 0.001$.}
\label{fig_sim}
\end{figure}

Based on the model introduced in the previous section, we performed numerical calculation.
The release of camphor molecules, described by Eq.~\eqref{supply}, was approximated using the expression:
\begin{equation}
f(\bm{r}, \bm{\ell}_j(t)) = \frac{1}{2 \pi \rho^2} \left[ 1 + \tanh \frac{ - \left( \left| \bm{r} - \bm{\ell}_j \right| - \rho \right)}{\delta} \right], \label{supply_smooth}
\end{equation}
 in order to reduce the effect of discretization, where $\delta$ is a positive constant for smoothing.

The parameters were set to be $\sigma = 0.001$, $\rho = 0.15$, and $\delta = 0.025$.
The distance between the disk center and the camphor pill center $\ell$, the number of camphor pills $N$, and the coefficient of the friction $\kappa$ were varied as parameters.
The time evolution was calculated with the Euler algorithm, and the diffusion was calculated with the explicit method.
Time step was $10^{-4}$ and the spatial mesh was $0.025$.
The force working on the camphor disk was calculated by summing the surface tension at 32 discrete points along the periphery.
To avoid the effect of the boundary, we calculated the camphor surface concentration up to 5 length units from the axis.
The Neumann conditions were applied at the boundaries.
The calculation started from the initial condition that $\phi = 1$ and $d\phi/dt = 0.1$.
The terminal angular velocity $\omega$ is set to be $d\phi/dt$ at $t = 100$, since we confirmed that the system was close to the stable stationary state at $t = 100$.

The numerical results are shown in Fig.~\ref{fig_sim}, in which we simultaneously plotted $\omega$ against $N$ for $\ell = 0.85, 1.35$, and $1.85$.
It should be noted that the ratios between the camphor pill radius and the disk radius are the same in experiments and numerical simulation.
In Fig.~\ref{fig_sim} (a), (b), and (c), we plotted the results on $\omega$ against $N$ for different values of $\kappa$.
The intersections of the plots with different $\ell$ changed depending on $\kappa$, and therefore we hope we can estimate $\kappa$ from the experimental results.
In this case, the plot in (b) is most close to the experimental results, and thus we can estimate $\kappa \simeq 0.01$.

\section{Theoretical analysis}

\begin{figure}
\begin{center}
\includegraphics{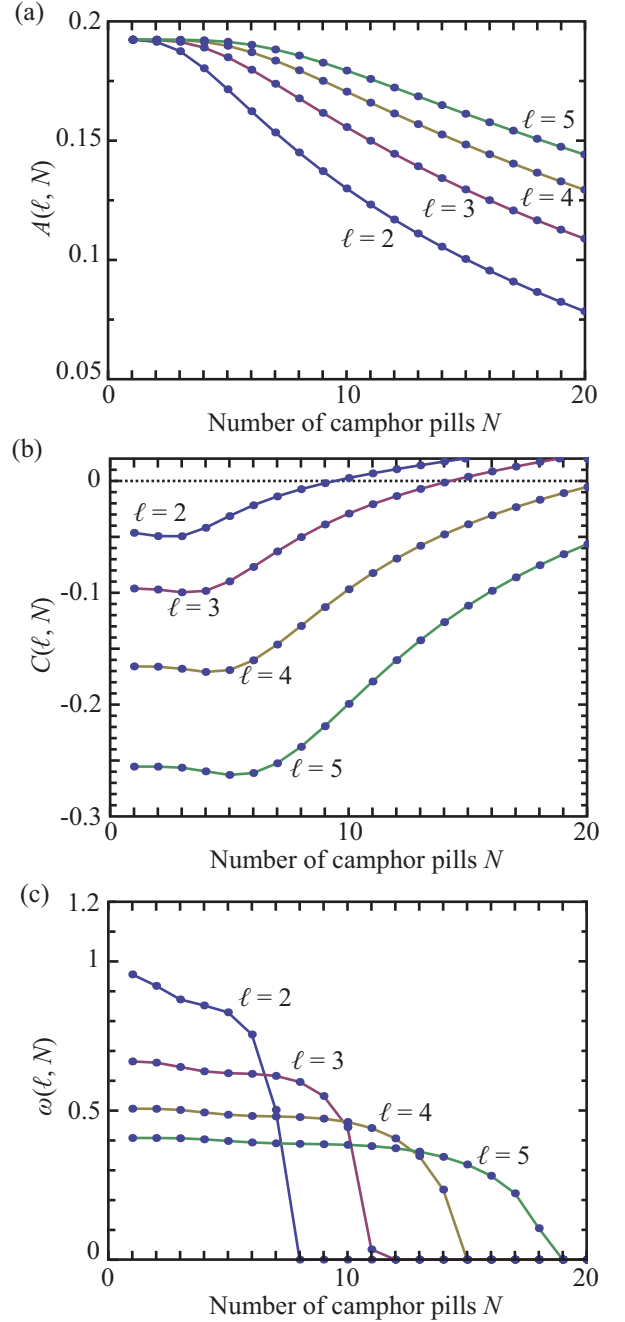}
\end{center}
\caption{Plots of (a) $A(\ell,N)$, (b) $C(\ell,N)$, and (c) $\omega(\ell,N)$ against $N$.
The distance $\ell$ is set to be $\ell =$ 2, 3, 4, and 5.
The radius of camphor disks $\rho$ is fixed to be 0.1.
The value of $\kappa$ is set to be $0.15$.}
\label{fig_theory}
\end{figure}

In this section, we derive the reduced evolution equation for the spinning of a disk with $N$ camphor pills.
We adopted Eq.~\eqref{diffusion} for the time evolution of concentration field, assuming that the support of function describing the release of camphor molecules is infinitesimally small:
\begin{equation}
f(\bm{r}, \bm{\ell}_j(t)) = \delta \left(\bm{r} - \bm{\ell}_j(t) \right), \label{supply-delta}
\end{equation}
instead of a finite size (cf. Eq.~\eqref{supply}).
Here $\delta(\cdot)$ is Dirac's delta function in a two-dimensional space.
The source term given by Eq.~\eqref{supply-delta} delivers the same amount of camphor as those described by Eq.~\eqref{supply}.

In our previous study~\cite{single_rotor,2d}, we derived the explicit form of the concentration field expanded with respect to the velocity, acceleration, jerk (i.e. the rate of change of acceleration), and so on of a camphor pill.
The concentration field at the position $\bm{r}$ originating from a camphor pill whose position is $\bm{\ell}(t)$ is described as:
\begin{align}
c_s(\bm{r}; \bm{\ell}) =& \frac{1}{2\pi} \mathcal{K}_0 \left ( d \right ) 
- \frac{1}{4\pi} \mathcal{K}_0 \left ( d \right ) \left [ \bm{d} \cdot \dot{\bm{\ell}} \right ] \nonumber \\
& + \frac{1}{16\pi} d \mathcal{K}_1 \left ( d \right ) \left [ \bm{d} \cdot \ddot{\bm{\ell}} \right ] - \frac{1}{16\pi} d \mathcal{K}_1 \left ( d \right ) |\dot{\bm{\ell}}|^2 \nonumber \\
&+ \frac{1}{16\pi} \mathcal{K}_0 \left ( d \right ) \left [ \bm{d} \cdot \dot{\bm{\ell}} \right ]^2 + \frac{1}{32\pi} d \mathcal{K}_1 \left ( d \right ) |\dot{\bm{\ell}}|^2 \left [ \bm{d} \cdot \dot{\bm{\ell}} \right ] \nonumber \\
&- \frac{1}{96 \pi} \mathcal{K}_0 \left ( d \right ) \left [ \bm{d} \cdot \dot{\bm{\ell}} \right ]^3 + \frac{1}{32 \pi} d^2 \mathcal{K}_2 \left ( d \right ) \left [ \dot{\bm{\ell}} \cdot \ddot{\bm{\ell}} \right ] \nonumber \\
&- \frac{1}{32\pi} d \mathcal{K}_1 \left ( d \right ) \left [ \bm{d} \cdot \dot{\bm{\ell}} \right ] \left [ \bm{d} \cdot \ddot{\bm{\ell}} \right ] \nonumber \\
&- \frac{1}{96 \pi} d^2 \mathcal{K}_2 \left ( d \right ) \left [ \bm{d} \cdot \dddot{\bm{\ell}} \right ] + \mbox{(higher order terms)}, \label{cs}
\end{align}
where $\bm{d} = \bm{r} - \bm{\ell}$ and $d = \left| \bm{d} \right|$. $\mathcal{K}_n$ is the second-kind modified Bessel function of $n$-th order, and a dot ($\dot{}$) means the time derivative.
In this expansion, we assume that the camphor pill speed is sufficiently small.

In the case of a disk with $N$ camphor pills, the concentration field is expressed by summing up the concentration field originating from each camphor pill since the evolution equation for concentration field is linear.
Thus, the concentration field made by the disk whose center is at the origin is given by
\begin{align}
c(\bm{r}) = \sum_{j = 0}^{N-1} c_s(\bm{r}; \bm{\ell}_j). \label{linearity}
\end{align}

Considering that the driving force originates from the imbalance of surface tension,
the driving force working on the $j$-th pill $\bm{F}_j$ can be calculated as follows:
\begin{align}
\frac{1}{\pi \rho^2} \bm{F}_j =& - \sum_{k = 0}^{N-1} \lim_{\rho \to +0} \frac {1}{\pi \rho^2} \int_{0}^{2 \pi} c_s(\bm{\ell}_j + \rho \bm{e}(\theta); \bm{\ell}_k) \bm{e}(\theta) \rho d\theta \nonumber \\
=& \frac{1}{4\pi} \left ( -\gamma +\log \frac{2}{\rho} \right ) \dot{\bm{\ell}}_j - \frac{1}{16\pi} \ddot{\bm{\ell}}_j - \frac{1}{32\pi} \left | \dot{\bm{\ell}}_j \right |^2 \dot{\bm{\ell}}_j \nonumber \\ &+ \frac{1}{48 \pi} \dddot{\bm{\ell}}_j - \sum_{k \neq j} \nabla c_s(\bm{\ell}_j; \bm{\ell}_k) + \mathcal{O}(\rho^1), \label{force_0}
\end{align}
where $\rho$ is considered to be an infinitesimally small parameter corresponding to the radius of a camphor pill.
It should be noted that $\nabla c_s (\bm{\ell}_j; \bm{\ell}_k) = \left . \nabla c_s (\bm{r}; \bm{\ell}_k) \right |_{\bm{r} = \bm{\ell}_j}$.

By explicitly calculating $\nabla c_s(\bm{\ell}_j; \bm{\ell}_k)$ and substituting the results into Eq.~\eqref{force_0}, the torque working on the $j$-th pill, $\mathcal{T}_j$, is obtained by taking the vector product of the radial vector and the force, 
\begin{equation}
 \mathcal{T}_j = \bm{\ell}_j \times \bm{F}_j.
\end{equation}
Since the torques acting on pills are identical due to the geometric symmetry, we finally obtain the total torque $\mathcal{T}$ working on the disk
\begin{equation}
\mathcal{T} = \sum_{j = 0}^{N-1} \mathcal{T}_j = N \mathcal{T}_0. \label{sumtorque}
\end{equation}
Therefore the reduced equation for the time evolution of the angle describing disk position $\phi(t)$ reads:
\begin{align}
\left ( \sigma + B(\ell,N) \right ) \frac{d^2 \phi}{dt^2}
= \left ( A(\ell,N) - \kappa \right ) \frac{d \phi}{dt} + C(\ell,N) \left( \frac{d \phi}{dt} \right)^3, \label{ee}
\end{align}
where $A(\ell,N)$, $B(\ell,N)$, and $C(\ell,N)$ are given as
\begin{align}
A(\ell,N) =& \frac{1}{4\pi} \left ( -\gamma_{\rm Euler} +\log \frac{2}{\rho} \right. \nonumber \\
&+ \left. \sum_{j=1}^{N-1} \left [ - \ell \mathcal{K}_1 \left ( 2 \ell \left | \sin \left (\frac{\pi j}{N} \right ) \right | \right ) \frac{\sin^2 \left ( \frac{2\pi j}{N} \right )}{2 \left | \sin \left ( \frac{\pi j}{N} \right ) \right |} \right. \right. \nonumber \\
& \left. \left. + \mathcal{K}_0 \left ( 2 \ell \left | \sin \left (\frac{\pi j}{N} \right ) \right | \right ) \cos \left ( \frac{2\pi j}{N} \right ) \right] \right), \label{A}
\end{align}
\begin{align}
& B(\ell, N) \nonumber \\
&= \frac{1}{16\pi} \left ( 1 - \sum_{j=1}^{N-1} \left [ \mathcal{K}_0 \left ( 2 \ell \left | \sin \left ( \frac{\pi j}{N} \right ) \right | \right ) \ell^2 \sin^2 \left ( \frac{2\pi j}{N} \right ) \right.\right. \nonumber \\
& \quad \left.\left.- 2 \mathcal{K}_1 \left ( 2 \ell \left | \sin \left ( \frac{\pi j}{N} \right ) \right | \right ) \ell \left | \sin \left ( \frac{\pi j}{N} \right ) \right | \cos \left ( \frac{2\pi j}{N} \right ) \right ] \right ), \label{B}
\end{align}
\begin{align}
& C(\ell, N) \nonumber \\
&= \frac{1}{192 \pi} \left ( -6 \ell^2 -4 \right. \nonumber \\
&+ \left. \sum_{j=1}^{N-1} \left [ 12 \mathcal{K}_0 \left ( 2 \ell \left | \sin \left ( \frac{\pi j}{N} \right ) \right | \right ) \ell^4 \sin^2 \left ( \frac{2\pi j}{N} \right ) \cos \left ( \frac{2\pi j}{N} \right ) \right.\right. \nonumber \\
& \left. \left. - \mathcal{K}_1 \left ( 2 \ell \left | \sin \left ( \frac{\pi j}{N} \right ) \right | \right ) \ell^5 \frac{\sin^4 \left ( \frac{2\pi j}{N} \right ) }{\left | \sin \left ( \frac{\pi j}{N} \right ) \right |} 
\right . \right . \nonumber \\
& \left. \left . + 4 \mathcal{K}_1 \left ( 2 \ell \left | \sin \left ( \frac{\pi j}{N} \right ) \right | \right ) \ell^3 \left | \sin \left ( \frac{\pi j}{N} \right ) \right | \right. \right. \nonumber \\
&\quad \times \left. \left. \left \{ - 3 \cos^2 \left ( \frac{2 \pi j}{N} \right ) + 4 \sin^2 \left ( \frac{2 \pi j}{N} \right ) \right \} \right.\right. \nonumber \\
& \left . \left . + 8 \mathcal{K}_2 \left ( 2 \ell \left | \sin \left ( \frac{\pi j}{N} \right ) \right | \right ) \ell^2 \sin^2 \left ( \frac{\pi j}{N} \right ) \cos \left ( \frac{2 \pi j}{N} \right ) \right ] \right ) , \label{C}
\end{align}
where $\gamma_{\rm Euler}$ is Euler-Mascheroni constant ($\simeq 0.577$).
In the process of calculation, we have neglected the term proportional to $d^3 \phi/dt^3$.
The detailed process of calculation is shown in Appendix \ref{detail}.
It should be noted that $B(\ell, N)$ is always positive, while $A(\ell, N)$ and $C(\ell, N)$ change their signs depending on the parameters.
The positive value of $B(\ell, N)$ means that the stationary solution and its stability does not change in the viscosity limit when $\sigma$ in Eq.~\eqref{ee} goes to zero.

At the bifurcation point, the coefficient of $d \phi/dt$ in the right side of Eq.~\eqref{ee} should be $0$, i.e., $\kappa = A(\ell,N)$.
The friction coefficient $\kappa$ depends on the water level, but in our experiments the water level was fixed so it can be regarded as a constant.
The coefficient $A(\ell,N)$ is plotted against $N$ in Fig.~\ref{fig_theory}(a) with $\ell =$ 2, 3, 4, and 5.
The coefficient $A(\ell,N)$ monotonically decreases with an increase in $N$ for every $\ell$.

If $A(\ell,N)$ is smaller than $\kappa$, then the disk stops.
Thus, a monotonical decrease in $A(\ell,N)$ with an increase in $N$ means that there is a critical number $N_c$ such that the disk stops for $N \geq N_c$.

The stable angular velocity $\omega(\ell, N)$ is given by $\omega(\ell, N) = \pm \sqrt{(A(\ell, N) - \kappa) / (-C(\ell,N))}$, if $A(\ell, N) - \kappa$ is positive and $C(\ell, N)$ is negative.
Figure \ref{fig_theory}(c) shows the plot of $\omega$ against $N$ with $\ell =$ 2, 3, 4, and 5.
When the coefficient $C(\ell, N)$ is positive, then the bifurcation is subcritical, and a higher order term (e.g. fifth-order term) is supposed to suppress the divergence of the angular velocity, which is out of scope of the present analysis.

\section{Discussion and Conclusion}

In the paper, we analyzed bifurcation between standing and rotating states of a disk propelled by a number of camphor pills.

Experiments on the stationary rotation of a disk located on the water surface were motivation for the present study.
The experiments revealed highly nonlinear behavior of the angular velocity $\omega (\ell,N)$ as a function of the disk size $\ell$ and of the number of pills propelling it $N$.
We observed that for a moderate number of pills ($ 5 \le N \le 12$) the angular velocity was a decreasing function of the disk size: larger disks rotated slower than smaller ones.
For a fixed disk radius and a moderate number of pills, the angular velocity $\omega (\ell ,N)$ was a slowly decreasing function of $N$.
However, if the number of pills approached a critical value $N_c$ then $\omega (\ell,N)$ rapidly dropped, and the disk did not rotate for all $N \ge N_c$.
The value of $N_c$ depended on the disk radius, and it was $14$, $22$ and $25$ for $ \ell = 0.85$, $1.35$ and $1.85$~cm, respectively.

We investigated the nonlinear behavior of the angular velocity of a disk driven by multiple camphor pills focusing on the bifurcation between rotation and rest.
It was reproduced by a mathematical model describing the dynamics of camphor surface concentration, its influence on the surface tension, and driving torque exerting on the disk through concentration-dependent surface tension.
The model we have used describes hydrodynamic effects and associated camphor transport via effective diffusion constant, that should be optimized for a particular system.
It uses scaled variables selected in such way that the effective diffusion constant $D$ as well as the combined rates of camphor evaporation and dissolution in water are equal to 1.
One of the adjustable model parameters is the friction coefficient $\kappa$.

\begin{figure}
\begin{center}
\includegraphics{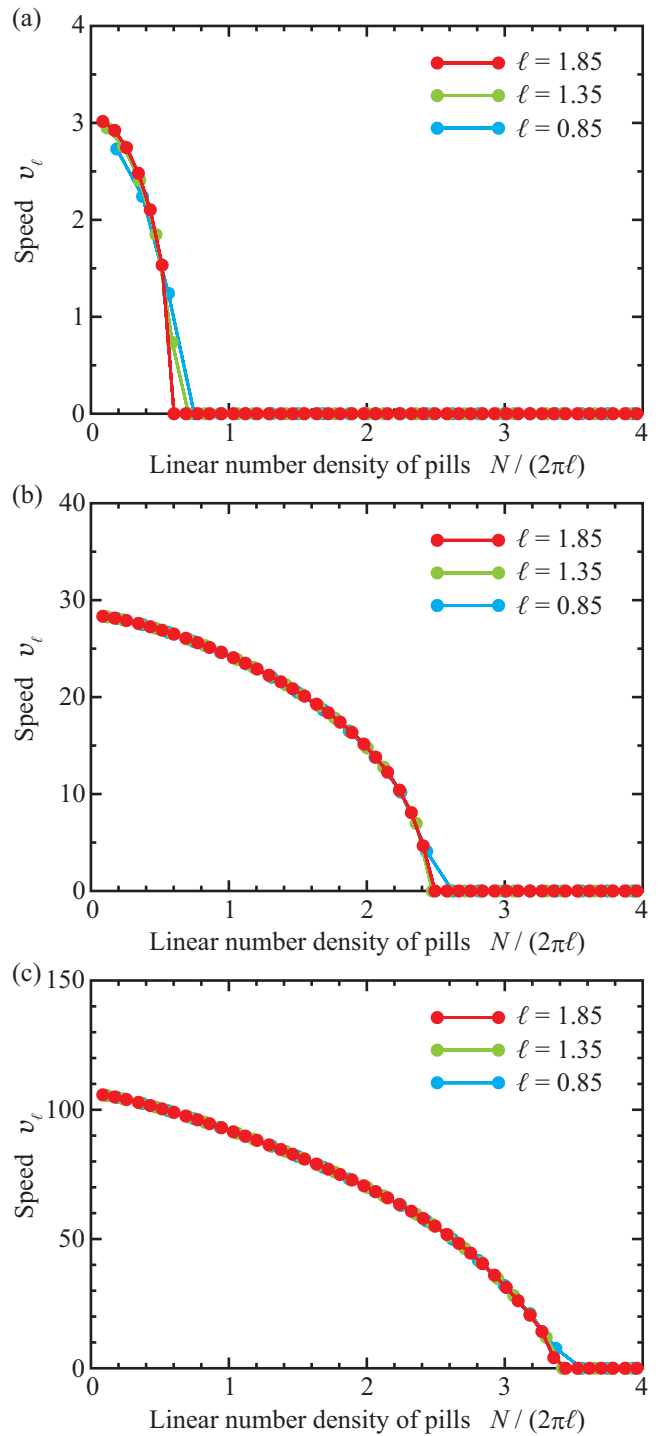}
\end{center}
\caption{Results of numerical calculation on the stationary speed, $v_{\ell} = \ell \omega(\ell, N)$, as a function of the linear number density of pills, $N/(2 \pi \ell)$.
Cyan, green, and red plots correspond to $\ell =0.85$, $1.35$, and $1.85$, respectively.
(a) $\kappa = 0.1$, (b) $\kappa = 0.01$, and (c) $\kappa = 0.001$.}
\label{fig_6}
\end{figure}

The numerical results for the speed of pills $v_\ell(\ell,N) = \ell \omega(\ell,N)$ as a function of the linear number density of pills $ N/(2 \pi \ell)$ are shown in Fig.~\ref{fig_6}.
It can be seen that, outside the bifurcation region, $v_\ell(\ell,N)$ was the same for different pairs ($\ell$, $N$) that lead to the same linear number density of pills.
The linear number density of pills is a local quantity.
It can be expected that the disk behavior solely depends on the density when the diameter of the disk is greater than the characteristic length of the diffusion (in our case it is equal to $1$) because the camphor pills hardly interact with the pills at the opposite side of the disk.
Such universal behavior of $v_\ell(\ell,N)$ observed for large disks, especially the density at which the transition between rotation and rest occurs, can be used to estimate the value of $\kappa$ that matches experimental results.

\begin{figure}
\begin{center}
\includegraphics{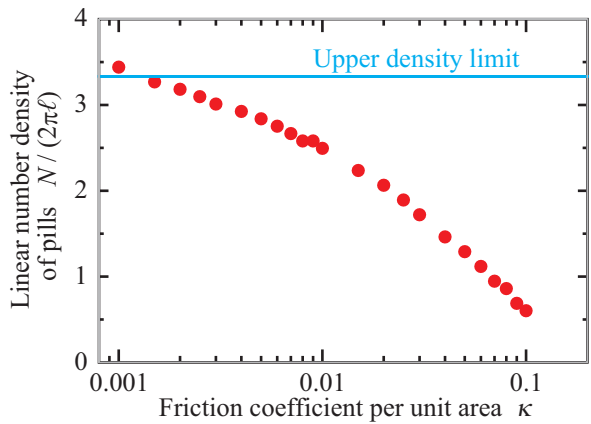}
\end{center}
\caption{Phase diagram illustrating different disk behavior depending on $\kappa$ and the linear number density of pills.
Dots mark pairs ($\kappa$, $N_c/(2 \pi \ell)$).
The dashed line represents the highest linear disk density for $\rho = 0.15$.
In all cases represented by points ($\kappa$, $N/(2 \pi \ell)$) below the dots, the disk rotates.
For the systems characterized by ($\kappa$, $N/(2 \pi \ell)$) above the dots, the disk does not move.
The numerical results for $\ell = 1.85$ were used for the present plot.}
\label{fig_7}
\end{figure}

Figure~\ref{fig_7} illustrates the phase diagram of disk behavior in the variables $\kappa$ and the linear number density of pills $N/(2 \pi \ell)$ calculated for $\rho = 0.15$. 
Dots mark pairs ($\kappa$, $N_c/(2 \pi \ell)$) corresponding to the smallest number of pills for which $v_\ell (\ell,N) = 0$.
They separate the phase space into regions where the disk rotates (below the dots) and where it does not (above the dots).

\begin{figure}
\begin{center}
\includegraphics{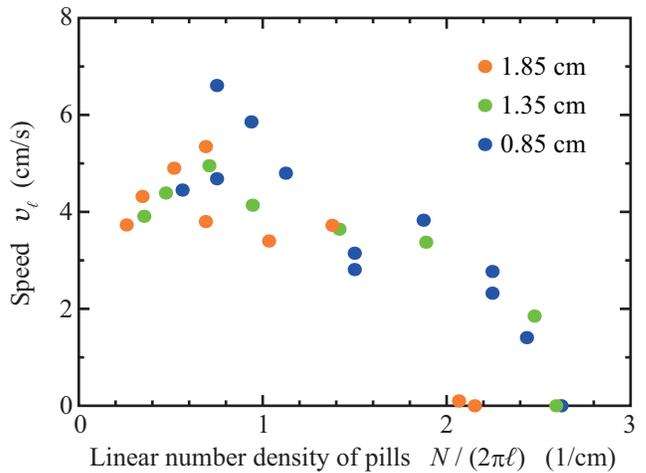}
\end{center}
\caption{Experimental results for $v_\ell$ as a function of $N/(2 \pi \ell)$.
Blue, green, and orange points correspond to $\ell =0.85$, $1.35$, and $1.85$~cm, respectively.}
\label{fig_8}
\end{figure}

The combined experimental results for $v_\ell (N/(2 \pi \ell))$ as a function of the linear number density of pills are shown in Fig.~\ref{fig_8}.
It can be seen that results for $\ell = 0.85$~cm and $\ell = 1.35$~cm nicely overlap, confirming universal behavior of $v_\ell (N/(2 \pi \ell))$.
The location of a critical point can be approximated by selecting $\kappa \sim 0.01$.
The majority of $v_\ell( N/(2 \pi \ell))$ values for $\ell = 1.85$~cm was smaller than those for disks of smaller radii.
Also the critical linear number density of pills for $\ell = 1.85$~cm was smaller than those for $\ell = 0.85$~cm or $\ell = 1.35$~cm.
Results shown in Fig.~\ref{fig_6} suggest that the value of $\kappa$ corresponding to $\ell = 1.85$~cm should be larger than $0.01$. The discrepancy can be explained by the fact that the largest disk was heavier and the pills were immersed deeper in water, and thus a larger friction coefficient should be applied to describe its motion.

We also discussed the model of disk rotation in the limit where camphor pills are infinitesimally small.
Within this assumption, we derived an analytical expression for camphor concentration profile and reduced equation of motion to the form that allows to find the bifurcation point and to estimate the stationary value of $\omega (\ell,N)$.
The results of the analytical model correctly reflect the basic experimental results: the value of $N_c$ increases with the disk radius and, outside the neighborhood of bifurcation, $\omega (\ell,N)$ is a decreasing function of $\ell$.
Also the shapes of $\omega(\ell,N)$ as a function of $N$ are similar to those seen in experiments (cf. Fig.~\ref{fig_exp_vel} and Fig.~\ref{fig_theory}(c)).

\begin{acknowledgments}

The authors are grateful to Professor S.~Nakata for his helpful comments. This work was supported by JSPS-PAN Bilateral Joint Research Program `` Spatio-temporal patterns of elements driven by self-generated, geometrically constrained flows'' between Japan and the Polish Academy of Sciences and by
the European Unions Horizon
2020 research and innovation programme under the
Marie Skodowska-Curie grant agreement No 734276 (N.A.) with additional support from the
Ministry of Science and Higher Education of Poland, agreement no 
3854/H2020/17/2018/2. Another author (Y.K.) is grateful for the support within
JSPS KAKENHI Grant Number JP17J05270 and the Cooperative Research Program of ``Network Joint Research Center for Materials and Devices'' No.~20181023.

\end{acknowledgments}

\appendix

\section{Detailed calculation \label{detail}}

In this section, we show the detailed derivation of Eq.~\eqref{ee} with Eqs.~\eqref{A} to \eqref{C}.

In Eq.~\eqref{force_0}, the last listed term represents the force originating from the camphor concentration released from the other pills, while the first four terms describe the force originating from the camphor concentration released from the considered pill itself.
Here we set
\begin{equation}
\bm{F}_j = \bm{F}_j^{\rm (self)} + \sum_{k \neq j} \bm{F}_{j,k}^{\rm (other)},
\end{equation}
where
\begin{align}
\frac{1}{\pi \rho^2} \bm{F}_j^{\rm (self)} =& \frac{1}{4\pi} \left ( -\gamma +\log \frac{2}{\rho} \right ) \dot{\bm{\ell}}_j - \frac{1}{16\pi} \ddot{\bm{\ell}}_j \nonumber \\ & - \frac{1}{32\pi} \left | \dot{\bm{\ell}}_j \right |^2 \dot{\bm{\ell}}_j + \frac{1}{48 \pi} \dddot{\bm{\ell}}_j + \mathcal{O}(\rho^1),
\end{align}
and
\begin{equation}
\frac{1}{\pi \rho^2} \bm{F}_{j,k}^{\rm (other)} = -\nabla c_s(\bm{\ell}_j; \bm{\ell}_k).
\end{equation}
By explicitly calculating the gradient of $c_s(\bm{r}; \bm{\ell})$ as \cite{single_rotor}
\begin{widetext}
\begin{align}
\nabla c_s(\bm{r}; \bm{\ell}) =& \frac{\partial}{\partial \bm{r}} c_s(\bm{r}; \bm{\ell}) \\
=& \frac{1}{2\pi} \mathcal{K}'_0 \left ( |\bm{r} - \bm{\ell}| \right ) \frac{\bm{r} - \bm{\ell}}{|\bm{r} - \bm{\ell}|} 
- \frac{1}{4\pi} \mathcal{K}'_0 \left ( |\bm{r} - \bm{\ell}| \right ) \left [ (\bm{r} - \bm{\ell}) \cdot \dot{\bm{\ell}} \right ] \frac{\bm{r} - \bm{\ell}}{|\bm{r} - \bm{\ell}|} 
- \frac{1}{4\pi} \mathcal{K}_0 \left ( |\bm{r} - \bm{\ell}| \right ) \dot{\bm{\ell}} \nonumber \\
& - \frac{1}{16\pi} \mathcal{K}_0 \left ( |\bm{r} - \bm{\ell}| \right ) \left [ (\bm{r} - \bm{\ell}) \cdot \ddot{\bm{\ell}} \right ] (\bm{r} - \bm{\ell}) 
+ \frac{1}{16\pi} |\bm{r} - \bm{\ell}| \mathcal{K}_1 \left ( |\bm{r} - \bm{\ell}| \right ) \ddot{\bm{\ell}}
+ \frac{1}{16 \pi} \mathcal{K}_0 \left ( |\bm{r} - \bm{\ell}| \right ) |\dot{\bm{\ell}}|^2 (\bm{r} - \bm{\ell}) \nonumber \\
& + \frac{1}{16\pi} \mathcal{K}'_0 \left ( |\bm{r} - \bm{\ell}| \right ) \left [ (\bm{r} - \bm{\ell}) \cdot \dot{\bm{\ell}} \right ]^2 \frac{\bm{r} - \bm{\ell}}{|\bm{r} - \bm{\ell}|} 
+ \frac{1}{8\pi} \mathcal{K}_0 \left ( |\bm{r} - \bm{\ell}| \right ) \left [ (\bm{r} - \bm{\ell}) \cdot \dot{\bm{\ell}} \right ] \dot{\bm{\ell}} \nonumber \\
& - \frac{1}{32\pi} \mathcal{K}_0 \left ( |\bm{r} - \bm{\ell}| \right ) |\dot{\bm{\ell}}|^2 \left [ (\bm{r} - \bm{\ell}) \cdot \dot{\bm{\ell}} \right ] (\bm{r} - \bm{\ell}) 
+ \frac{1}{32\pi} |\bm{r} - \bm{\ell}| \mathcal{K}_1 \left ( |\bm{r} - \bm{\ell}| \right ) |\dot{\bm{\ell}}|^2 \dot{\bm{\ell}} \nonumber \\
& - \frac{1}{96\pi} \mathcal{K}'_0 \left ( |\bm{r} - \bm{\ell}| \right ) \left [ (\bm{r} - \bm{\ell}) \cdot \dot{\bm{\ell}} \right ]^3 \frac{\bm{r} - \bm{\ell}}{|\bm{r} - \bm{\ell}|}
- \frac{1}{32\pi} \mathcal{K}_0 \left ( |\bm{r} - \bm{\ell}| \right ) \left [ (\bm{r} - \bm{\ell}) \cdot \dot{\bm{\ell}} \right ]^2 \dot{\bm{\ell}} \nonumber \\
& - \frac{1}{32\pi} |\bm{r} - \bm{\ell}| \mathcal{K}_1 \left ( |\bm{r} - \bm{\ell}| \right ) \left ( \dot{\bm{\ell}} \cdot \ddot{\bm{\ell}} \right ) (\bm{r} - \bm{\ell})
+ \frac{1}{32\pi} \mathcal{K}_0 \left ( |\bm{r} - \bm{\ell}| \right ) \left [ (\bm{r} - \bm{\ell}) \cdot \dot{\bm{\ell}} \right ] \left [ (\bm{r} - \bm{\ell}) \cdot \ddot{\bm{\ell}} \right ] (\bm{r} - \bm{\ell}) \nonumber \\
& - \frac{1}{32\pi} |\bm{r} - \bm{\ell}| \mathcal{K}_1 \left ( |\bm{r} - \bm{\ell}| \right ) \left [ (\bm{r} - \bm{\ell}) \cdot \ddot{\bm{\ell}} \right ] \dot{\bm{\ell}}
- \frac{1}{32\pi} |\bm{r} - \bm{\ell}| \mathcal{K}_1 \left ( |\bm{r} - \bm{\ell}| \right ) \left [ (\bm{r} - \bm{\ell}) \cdot \dot{\bm{\ell}} \right ] \ddot{\bm{\ell}} \nonumber \\
& + \frac{1}{96\pi} |\bm{r} - \bm{\ell}| \mathcal{K}_1 \left ( |\bm{r} - \bm{\ell}| \right ) \left [ (\bm{r} - \bm{\ell}) \cdot \dddot{\bm{\ell}} \right ] (\bm{r} - \bm{\ell})
- \frac{1}{96\pi} |\bm{r} - \bm{\ell}|^2 \mathcal{K}_2 \left ( |\bm{r} - \bm{\ell}| \right ) \dddot{\bm{\ell}}, \label{A5}
\end{align}
\end{widetext}
we can obtain the explicit form of $\bm{F}_{j,k}^{\rm (other)}$.
Here we used the property of modified Bessel function \cite{Watson}:\\
$z {\mathcal{K}_\nu}'(z) + \nu \mathcal{K}_\nu (z) = -z \mathcal{K}_{\nu-1}(z)$.

The torque $\mathcal{T}_{j,j}$ working on the $j$-th camphor pill originating from $\bm{F}_j^{\rm (self)}$, is given by
\begin{widetext}
\begin{align}
\frac{1}{\pi \rho^2} \mathcal{T}_{j,j} =
& \lim_{\bm{r} \to \bm{\ell}_0} \left [ \frac{1}{4\pi} \mathcal{K}_0 (|\bm{r} - \bm{\ell}_0| ) \left ( \bm{\ell}_0 \times \dot{\bm{\ell}}_0 \right ) 
- \frac{1}{16\pi} |\bm{r} - \bm{\ell}| \mathcal{K}_1 \left ( |\bm{r} - \bm{\ell}_0| \right ) \left ( \bm{\ell}_0 \times \ddot{\bm{\ell}}_0 \right ) \right .
\nonumber \\
& \left . - \frac{1}{32\pi} |\bm{r} - \bm{\ell}_0| \mathcal{K}_1 \left ( |\bm{r} - \bm{\ell}_0| \right ) |\dot{\bm{\ell}}_0|^2 \left ( \bm{\ell}_0 \times\dot{\bm{\ell}}_0 \right ) 
+ \frac{1}{96\pi} |\bm{r} - \bm{\ell}_0|^2 \mathcal{K}_2 \left ( |\bm{r} - \bm{\ell}_0| \right ) \left ( \bm{\ell}_0 \times \dddot{\bm{\ell}}_0 \right ) \right ] \nonumber \\
=& \frac{1}{4\pi} \left ( - \gamma_{\mathrm{Euler}} + \log\frac{2}{\epsilon} \right )\ell^2 \dot{\phi} - \frac{1}{16\pi} \ell^2 \ddot{\phi} - \frac{1}{32\pi} \ell^4 \dot{\phi}^3 + \frac{1}{48\pi} \ell^2 (\dddot{\phi} - \dot{\phi}^3).
\end{align}
\end{widetext}
Here we used \cite{Watson} $\lim_{x \to +0}\mathcal{K}_0 (x) = - \gamma_{\mathrm{Euler}} + \log (2/x)$, $\lim_{x \to +0} x \mathcal{K}_1 (x) = 1$, $\lim_{x \to +0} x^2 \mathcal{K}_2 (x) = 2$.

From Eq.~\eqref{sumtorque}, we only have to obtain $\mathcal{T}_0$ considering the system symmetry, and it is calculated as:
\begin{align}
\mathcal{T}_0 =& \bm{\ell}_0 \times \bm{F}_0 \nonumber \\
=& \bm{\ell}_0 \times \bm{F}_0^{\rm (self)} + \bm{\ell}_0 \times \sum_{k = 1}^{N-1} \bm{F}_{0,k}^{\rm (other)} \nonumber\\
=& \mathcal{T}_{0,0} + \sum_{k = 1}^{N-1} \mathcal{T}_{0,k}.
\end{align}

Therefore, $\mathcal{T}_{0,k}$ is obtained from Eq.~\eqref{A5} as 
\begin{widetext}
\begin{align}
\frac{1}{\pi \rho^2} \mathcal{T}_{0,k} =& - [ \bm{\ell}_0 \times \nabla c_s(\bm{\ell}_0; \bm{\ell}_k) ] \nonumber \\
=& \frac{1}{2\pi} \mathcal{K}'_0 \left ( |\bm{\ell}_0 - \bm{\ell}_k| \right ) \frac{\bm{\ell}_0 \times \bm{\ell}_k}{|\bm{\ell}_0 - \bm{\ell}_k|} 
- \frac{1}{4\pi} \mathcal{K}'_0 \left ( |\bm{\ell}_0 - \bm{\ell}_k| \right ) \left ( \bm{\ell}_0 \cdot \dot{\bm{\ell}}_k \right ) \frac{\bm{\ell}_0 \times \bm{\ell}_k}{|\bm{\ell}_0 - \bm{\ell}_k|} 
+ \frac{1}{4\pi} \mathcal{K}_0 \left ( |\bm{\ell}_0 - \bm{\ell}_k| \right ) \left ( \bm{\ell}_0 \times \dot{\bm{\ell}}_k \right ) \nonumber \\
& - \frac{1}{16\pi} \mathcal{K}_0 \left ( |\bm{\ell}_0 - \bm{\ell}_k| \right ) \left [ (\bm{\ell}_0 - \bm{\ell}_k) \cdot \ddot{\bm{\ell}}_k \right ] (\bm{\ell}_0 \times \bm{\ell}_k) 
- \frac{1}{16\pi} |\bm{\ell}_0 - \bm{\ell}_k| \mathcal{K}_1 \left ( |\bm{\ell}_0 - \bm{\ell}_k| \right ) \left ( \bm{\ell}_0 \times \ddot{\bm{\ell}}_k \right )
\nonumber \\
&+ \frac{1}{16 \pi} \mathcal{K}_0 \left ( |\bm{\ell}_0 - \bm{\ell}_k| \right ) |\dot{\bm{\ell}}_k|^2 (\bm{\ell}_0 \times \bm{\ell}_k) + \frac{1}{16\pi} \mathcal{K}'_0 \left ( |\bm{\ell}_0 - \bm{\ell}_k| \right ) \left ( \bm{\ell}_0 \cdot \dot{\bm{\ell}}_k \right )^2 \frac{\bm{\ell}_0 \times \bm{\ell}_k}{|\bm{\ell}_0 - \bm{\ell}_k|} \nonumber \\
&- \frac{1}{8\pi} \mathcal{K}_0 \left ( |\bm{\ell}_0 - \bm{\ell}_k| \right ) \left ( \bm{\ell}_0 \cdot \dot{\bm{\ell}}_k \right ) \left ( \bm{\ell}_0 \times \dot{\bm{\ell}}_k \right ) - \frac{1}{32\pi} \mathcal{K}_0 \left ( |\bm{\ell}_0 - \bm{\ell}_k| \right ) |\dot{\bm{\ell}}_k|^2 \left ( \bm{\ell}_0 \cdot \dot{\bm{\ell}}_k \right )(\bm{\ell}_0 \times \bm{\ell}_k) 
\nonumber \\
&
- \frac{1}{32\pi} |\bm{\ell}_0 - \bm{\ell}_k| \mathcal{K}_1 \left ( |\bm{\ell}_0 - \bm{\ell}_k| \right ) |\dot{\bm{\ell}}_k|^2 (\bm{\ell}_0 \times \dot{\bm{\ell}}_k ) - \frac{1}{96\pi} \mathcal{K}'_0 \left ( |\bm{\ell}_0 - \bm{\ell}_k| \right ) \left ( \bm{\ell}_0 \cdot \dot{\bm{\ell}}_k \right )^3 \frac{\bm{\ell}_0 \times \bm{\ell}_k}{|\bm{\ell}_0 - \bm{\ell}_k|}
\nonumber \\
&
+ \frac{1}{32\pi} \mathcal{K}_0 \left ( |\bm{\ell}_0 - \bm{\ell}_k| \right ) \left ( \bm{\ell}_0 \cdot \dot{\bm{\ell}}_k \right )^2 \left ( \bm{\ell}_0 \times \dot{\bm{\ell}}_k \right ) - \frac{1}{32\pi} |\bm{\ell}_0 - \bm{\ell}_k| \mathcal{K}_1 \left ( |\bm{\ell}_0 - \bm{\ell}_k| \right ) \left ( \dot{\bm{\ell}}_k \cdot \ddot{\bm{\ell}}_k \right ) (\bm{\ell}_0 \times \bm{\ell}_k)
\nonumber \\
&
+ \frac{1}{32\pi} \mathcal{K}_0 \left ( |\bm{\ell}_0 - \bm{\ell}_k| \right ) \left ( \bm{\ell}_0 \cdot \dot{\bm{\ell}}_k \right ) \left [ (\bm{\ell}_0 - \bm{\ell}_k) \cdot \ddot{\bm{\ell}}_k \right ] (\bm{\ell}_0 \times \bm{\ell}_k) 
\nonumber \\
& + \frac{1}{32\pi} |\bm{\ell}_0 - \bm{\ell}_k| \mathcal{K}_1 \left ( |\bm{\ell}_0 - \bm{\ell}_k| \right ) \left [ (\bm{\ell}_0 - \bm{\ell}_k) \cdot \ddot{\bm{\ell}}_k \right ] \left ( \bm{\ell}_0 \times \dot{\bm{\ell}}_k \right )
+ \frac{1}{32\pi} |\bm{\ell}_0 - \bm{\ell}_k| \mathcal{K}_1 \left ( |\bm{\ell}_0 - \bm{\ell}_k| \right ) \left ( \bm{\ell}_0 \cdot \dot{\bm{\ell}}_k \right ) \left ( \bm{\ell}_0 \times \ddot{\bm{\ell}}_k \right ) \nonumber \\
& + \frac{1}{96\pi} |\bm{\ell}_0 - \bm{\ell}_k| \mathcal{K}_1 \left ( |\bm{\ell}_0 - \bm{\ell}_k| \right ) \left [ (\bm{\ell}_0 - \bm{\ell}_k) \cdot \dddot{\bm{\ell}}_k \right ] (\bm{\ell}_0 \times \bm{\ell}_k)
+ \frac{1}{96\pi} |\bm{\ell}_0 - \bm{\ell}_k|^2 \mathcal{K}_2 \left ( |\bm{\ell}_0 - \bm{\ell}_k| \right ) \left ( \bm{\ell}_0 \times \dddot{\bm{\ell}}_k \right ) \nonumber \\
=& \frac{1}{4\pi} \mathcal{K}'_0 \left ( 2 \ell \left | \sin \left ( \frac{\pi k}{N} \right )\right | \right ) \frac{\ell \sin \left ( \frac{2\pi k}{N} \right ) }{\left | \sin \left ( \frac{\pi k}{N} \right ) \right |} + \frac{1}{8\pi} \mathcal{K}'_0 \left ( 2 \ell \left | \sin \left (\frac{\pi k}{N} \right ) \right | \right ) \ell^3 \frac{\sin^2 \left ( \frac{2\pi k}{N} \right )}{\left | \sin \left ( \frac{\pi k}{N} \right ) \right |} \dot{\phi} \nonumber \\
& + \frac{1}{4\pi} \mathcal{K}_0 \left ( 2 \ell \left | \sin \left ( \frac{\pi k}{N} \right ) \right | \right ) \ell^2 \cos \left ( \frac{2\pi k}{N} \right ) \dot{\phi} \nonumber \\
&
+ \frac{1}{16\pi} \mathcal{K}_0 \left ( 2 \ell \left | \sin \left ( \frac{\pi k}{N} \right ) \right | \right ) \ell^4 \sin \left ( \frac{2\pi k}{N} \right ) \left ( \dot{\phi}^2 \left ( \cos \frac{2\pi k}{N} - 1 \right ) + \ddot{\phi} \sin \frac{2\pi k}{N} \right ) \nonumber \\
& - \frac{1}{8\pi} \mathcal{K}_1 \left ( 2 \ell \left | \sin \left ( \frac{\pi k}{N} \right ) \right | \right ) \ell^3 \left | \sin \left ( \frac{\pi k}{N} \right ) \right | \left ( \ddot{\phi} \cos \left ( \frac{2\pi k}{N} \right ) - \dot{\phi}^2 \sin \left ( \frac{2\pi k}{N} \right ) \right ) \nonumber \\
&
+ \frac{1}{16\pi} \mathcal{K}_0 \left ( 2 \ell \left | \sin \left ( \frac{\pi k}{N} \right ) \right | \right ) \ell^4 \sin \left ( \frac{2\pi k}{N} \right ) \dot{\phi}^2 \nonumber \\
& + \frac{1}{32\pi} \mathcal{K}'_0 \left ( 2 \ell \left | \sin \left ( \frac{\pi k}{N} \right ) \right | \right ) \ell^5 \frac{\sin^3 \left ( \frac{2\pi k}{N} \right ) }{\left | \sin \left ( \frac{\pi k}{N} \right ) \right |} \dot{\phi}^2
+ \frac{1}{8\pi} \mathcal{K}_0 \left ( 2 \ell \left | \sin \left ( \frac{\pi k}{N} \right ) \right | \right ) \ell^4 \sin \left ( \frac{2\pi k}{N} \right ) \cos \left ( \frac{2\pi k}{N} \right ) \dot{\phi}^2 \nonumber \\
& + \frac{1}{32\pi} \mathcal{K}_0 \left ( 2 \ell \left | \sin \left ( \frac{\pi k}{N} \right ) \right | \right ) \ell^6 \sin^2 \left ( \frac{2\pi k}{N} \right ) \dot{\phi}^3 
- \frac{1}{16\pi} \mathcal{K}_1 \left ( 2 \ell \left | \sin \left ( \frac{\pi k}{N} \right ) \right | \right ) \ell^5 \left | \sin \left ( \frac{\pi k}{N} \right ) \right | \cos \left ( \frac{2 \pi k}{N} \right ) \dot{\phi}^3 \nonumber \\
&+ \frac{1}{192\pi} \mathcal{K}'_0 \left ( 2 \ell \left | \sin \left ( \frac{\pi k}{N} \right ) \right | \right ) \ell^7 \frac{\sin^4 \left ( \frac{2\pi k}{N} \right ) }{\left | \sin \left ( \frac{\pi k}{N} \right ) \right |} \dot{\phi}^3
+ \frac{1}{32\pi} \mathcal{K}_0 \left ( 2 \ell \left | \sin \left ( \frac{\pi k}{N} \right ) \right | \right ) \ell^6 \sin^2 \left ( \frac{2\pi k}{N} \right ) \cos \left ( \frac{2\pi k}{N} \right ) \dot{\phi}^3 \nonumber \\
& - \frac{1}{16\pi} \mathcal{K}_1 \left ( 2 \ell \left | \sin \left ( \frac{\pi k}{N} \right ) \right | \right ) \ell^5 \left | \sin \left ( \frac{\pi k}{N} \right ) \right | \sin \left ( \frac{2 \pi k}{N} \right ) \dot{\phi} \ddot{\phi} \nonumber \\
& + \frac{1}{32\pi} \mathcal{K}_0 \left ( 2 \ell \left | \sin \left ( \frac{\pi k}{N} \right ) \right | \right ) \ell^6 \sin^2 \left ( \frac{2\pi k}{N} \right )\left [ \dot{\phi}^2 \left ( \cos \left ( \frac{2\pi k}{N} \right ) - 1 \right ) + \ddot{\phi} \sin \left ( \frac{2\pi k}{N} \right ) \right ] \dot{\phi} \nonumber \\
& - \frac{1}{16\pi} \mathcal{K}_1 \left ( 2 \ell \left | \sin \left ( \frac{\pi k}{N} \right ) \right | \right ) \ell^5 \left | \sin \left ( \frac{\pi k}{N} \right ) \right | \cos \left ( \frac{2 \pi k}{N} \right ) \left [ \dot{\phi}^2 \left ( \cos \left ( \frac{2\pi k}{N} \right ) - 1 \right ) + \ddot{\phi} \sin\left ( \frac{2\pi k}{N} \right ) \right ] \dot{\phi} \nonumber \\
& + \frac{1}{16\pi} \mathcal{K}_1 \left ( 2 \ell \left | \sin \left ( \frac{\pi k}{N} \right ) \right | \right ) \ell^5 \left | \sin \left ( \frac{\pi k}{N} \right ) \right | \sin \left ( \frac{2 \pi k}{N} \right ) \left [ \dot{\phi}^2 \sin \left ( \frac{2\pi k}{N} \right ) - \ddot{\phi} \cos \left ( \frac{2\pi k}{N} \right )\right ] \dot{\phi} \nonumber \\
& - \frac{1}{48 \pi} \mathcal{K}_1 \left ( 2 \ell \left | \sin \left ( \frac{\pi k}{N} \right ) \right | \right ) \ell^5 \left | \sin \left ( \frac{\pi k}{N} \right ) \right | \sin \left ( \frac{2 \pi k}{N} \right ) \left [ (\dddot{\phi} - \dot{\phi}^3) \sin \left ( \frac{2 \pi k}{N} \right ) + 3 \dot{\phi} \ddot{\phi}\left ( \cos \left ( \frac{2 \pi k}{N} \right ) - 1 \right ) \right ] \nonumber \\
& + \frac{1}{24\pi} \mathcal{K}_2 \left ( 2 \ell \left | \sin \left ( \frac{\pi k}{N} \right ) \right | \right ) \ell^4 \sin^2 \left ( \frac{\pi k}{N} \right ) \left [ (\dddot{\phi} - \dot{\phi}^3) \cos \left ( \frac{2 \pi k}{N} \right ) - 3 \dot{\phi} \ddot{\phi} \sin \left ( \frac{2 \pi k}{N} \right ) \right ].
\end{align}
\end{widetext}
Here, we used the time derivative of Eq.~\eqref{position}.

We thus have the expression for the torque $\mathcal{T}_0$:
\begin{widetext}
\begin{align}
\frac{1}{\pi \rho^2} \mathcal{T}_{0} =& \frac{1}{\pi \rho^2} \mathcal{T}_{0,0} + \sum_{k=1}^{N-1} \frac{1}{\pi \rho^2} \mathcal{T}_{0,k} \nonumber \\
=& \frac{\ell^2}{4\pi} \left ( -\gamma +\log \frac{2}{\rho} + \sum_{k=1}^{N-1} \left [ - \ell \mathcal{K}_1 \left ( 2 \ell \left | \sin \left (\frac{\pi k}{N} \right ) \right | \right ) \frac{\sin^2 \left ( \frac{2\pi k}{N} \right )}{2 \left | \sin \left ( \frac{\pi k}{N} \right ) \right |} 
+ \mathcal{K}_0 \left ( 2 \ell \left | \sin \left (\frac{\pi k}{N} \right ) \right | \right ) \cos \left ( \frac{2\pi k}{N} \right ) \right ] \right ) \dot{\phi} \nonumber \\
& + \frac{\ell^2}{16\pi} \left ( - 1 + \sum_{k=1}^{N-1} \left [ \mathcal{K}_0 \left ( 2 \ell \left | \sin \left ( \frac{\pi k}{N} \right ) \right | \right ) \ell^2 \sin^2 \left ( \frac{2\pi k}{N} \right ) 
\right . \right. \nonumber \\
& \qquad \qquad \left. \left.
- 2 \mathcal{K}_1 \left ( 2 \ell \left | \sin \left ( \frac{\pi k}{N} \right ) \right | \right ) \ell \left | \sin \left ( \frac{\pi k}{N} \right ) \right | \cos \left ( \frac{2\pi k}{N} \right ) \right ] \right ) \ddot{\phi} \nonumber \\
& + \frac{\ell^2}{192 \pi} \left ( -6 \ell^2 -4 + \sum_{k=1}^{N-1} \left [ 12 \mathcal{K}_0 \left ( 2 \ell \left | \sin \left ( \frac{\pi k}{N} \right ) \right | \right ) \ell^4 \sin^2 \left ( \frac{2\pi k}{N} \right ) \cos \left ( \frac{2\pi k}{N} \right )
\right . \right. \nonumber \\
& \qquad \qquad \left. \left. - \mathcal{K}_1 \left ( 2 \ell \left | \sin \left ( \frac{\pi k}{N} \right ) \right | \right ) \ell^5 \frac{\sin^4 \left ( \frac{2\pi k}{N} \right ) }{\left | \sin \left ( \frac{\pi k}{N} \right ) \right |} 
\right . \right . \nonumber \\
& \qquad \qquad + 4 \mathcal{K}_1 \left ( 2 \ell \left | \sin \left ( \frac{\pi k}{N} \right ) \right | \right ) \ell^3 \left | \sin \left ( \frac{\pi k}{N} \right ) \right | \left \{ - 3 \cos^2 \left ( \frac{2 \pi k}{N} \right ) + 4 \sin^2 \left ( \frac{2 \pi k}{N} \right ) \right \} \nonumber \\
& \qquad \qquad \left . \left . + 8 \mathcal{K}_2 \left ( 2 \ell \left | \sin \left ( \frac{\pi k}{N} \right ) \right | \right ) \ell^2 \sin^2 \left ( \frac{\pi k}{N} \right ) \cos \left ( \frac{2 \pi k}{N} \right ) \right ] \right ) \dot{\phi}^3 \nonumber \\
& + \frac{\ell^2}{48 \pi} \left ( 1 + \sum_{k=1}^{N-1} \left [ - \mathcal{K}_1 \left ( 2 \ell \left | \sin \left ( \frac{\pi k}{N} \right ) \right | \right ) \ell^3 \left | \sin \left ( \frac{\pi k}{N} \right ) \right | \sin^2 \left ( \frac{2 \pi k}{N} \right ) \right.\right. \nonumber \\
& \qquad \qquad\left. \left.
+ 2 \mathcal{K}_2 \left ( 2 \ell \left | \sin \left ( \frac{\pi k}{N} \right ) \right | \right ) \ell^2 \sin^2 \left ( \frac{\pi k}{N} \right ) \cos \left ( \frac{2 \pi k}{N} \right ) \right ] \right ) \dddot{\phi}. \label{torque}
\end{align}
\end{widetext}
Reducing the number of terms we used:
\begin{align}
&\sum_{k=1}^{N-1} f\left (\left | \sin \left (\frac{\pi k}{N} \right ) \right | \right ) \sin \left ( \frac{2\pi k}{N} \right ) = 0, \\
& \sum_{k=1}^{N-1} f\left (\left | \sin \left (\frac{\pi k}{N} \right ) \right | \right ) \sin^3 \left ( \frac{2\pi k}{N} \right ) = 0, \\ 
&\sum_{k=1}^{N-1} f\left (\left | \sin \left (\frac{\pi k}{N} \right ) \right | \right ) \sin \left ( \frac{4\pi k}{N} \right ) = 0.
\end{align}

Thus, the equation for the position of the disk reads:
\begin{widetext}
\begin{align}
& N \sigma \ell^2 \ddot{\phi} \nonumber \\
=& - \kappa N \ell^2 \dot{\phi} 
 + \frac{N \ell^2 \dot{\phi}}{4\pi} \left ( -\gamma +\log \frac{2}{\rho} + \sum_{k=1}^{N-1} \left [ - \ell \mathcal{K}_1 \left ( 2 \ell \left | \sin \left (\frac{\pi k}{N} \right ) \right | \right ) \frac{\sin^2 \left ( \frac{2\pi k}{N} \right )}{2 \left | \sin \left ( \frac{\pi k}{N} \right ) \right |} + \mathcal{K}_0 \left ( 2 \ell \left | \sin \left (\frac{\pi k}{N} \right ) \right | \right ) \cos \left ( \frac{2\pi k}{N} \right ) \right ] \right ) \nonumber \\
& + \frac{N \ell^2}{16\pi} \left ( - 1 + \sum_{k=1}^{N-1} \left [ \mathcal{K}_0 \left ( 2 \ell \left | \sin \left ( \frac{\pi k}{N} \right ) \right | \right ) \ell^2 \sin^2 \left ( \frac{2\pi k}{N} \right ) 
- 2 \mathcal{K}_1 \left ( 2 \ell \left | \sin \left ( \frac{\pi k}{N} \right ) \right | \right ) \ell \left | \sin \left ( \frac{\pi k}{N} \right ) \right | \cos \left ( \frac{2\pi k}{N} \right ) \right ] \right ) \ddot{\phi} \nonumber \\
& + \frac{N \ell^2}{192 \pi} \left ( -6 \ell^2 -4 + \sum_{k=1}^{N-1} \left [ 12 \mathcal{K}_0 \left ( 2 \ell \left | \sin \left ( \frac{\pi k}{N} \right ) \right | \right ) \ell^4 \sin^2 \left ( \frac{2\pi k}{N} \right ) \cos \left ( \frac{2\pi k}{N} \right )
- \mathcal{K}_1 \left ( 2 \ell \left | \sin \left ( \frac{\pi k}{N} \right ) \right | \right ) \ell^5 \frac{\sin^4 \left ( \frac{2\pi k}{N} \right ) }{\left | \sin \left ( \frac{\pi k}{N} \right ) \right |} 
\right . \right . \nonumber \\
& \qquad \qquad + 4 \mathcal{K}_1 \left ( 2 \ell \left | \sin \left ( \frac{\pi k}{N} \right ) \right | \right ) \ell^3 \left | \sin \left ( \frac{\pi k}{N} \right ) \right | \left \{ - 3 \cos^2 \left ( \frac{2 \pi k}{N} \right ) + 4 \sin^2 \left ( \frac{2 \pi k}{N} \right ) \right \} \nonumber \\
& \qquad \qquad \left . \left . + 8 \mathcal{K}_2 \left ( 2 \ell \left | \sin \left ( \frac{\pi k}{N} \right ) \right | \right ) \ell^2 \sin^2 \left ( \frac{\pi k}{N} \right ) \cos \left ( \frac{2 \pi k}{N} \right ) \right ] \right ) \dot{\phi}^3 \label{rdc_eq.m}.
\end{align}
\end{widetext}
Here we neglect the term proportional to $\dddot{\phi}$.
By dividing the both sides of Eq.~\eqref{rdc_eq.m} with $N \ell^2$, we obtain Eqs.~\eqref{ee} to \eqref{C}.


\begin{thebibliography}{99}
\bibitem{footnote1} The frictional force working on the $j$-th camphor pill is represented as $-\pi \rho^2 \kappa (d\bm{\ell}_j / dt)$. Therefore the torque originating from this force is $-\pi \rho^2 \kappa \bm{\ell}_j \times (d\bm{\ell}_j / dt) = -\pi \rho^2 \kappa \ell^2 (d\phi/dt)$. By summing up the torque on $N$ particles, we obtain Eq.~\eqref{friction}.
\end{thebibliography}
\end{document}